\documentclass[utf8]{FrontiersinHarvard} 

\usepackage{url,hyperref,lineno,microtype,subcaption}
\usepackage[onehalfspacing]{setspace}
\usepackage{ulem}
\usepackage{hyperref}

\def\keyFont{\fontsize{8}{11}\helveticabold }
\def\firstAuthorLast{Yakovlev {et~al.}} 
\def\Authors{
Oleg Ya. Yakovlev\,$^{1,2*}$, 
Azamat F. Valeev\,$^{1,3,4}$,
Gennady G. Valyavin\,$^{1}$, 
Alexander V. Tavrov\,$^{2,5}$, 
Vitaly Aitov \,$^{1}$,
Guram Mitiani\,$^{1}$,
Oleg Korablev\,$^{2}$,
Gazinur Galazutdinov\,$^{4,1}$,
Grigory Beskin\,$^{1}$,
Eduard Emelianov\,$^{1}$,
Timur Fatkhullin\,$^{1}$,
Valery Vlasyuk\,$^{1}$,
Vyacheslav Sasyuk\,$^{6}$,
Alexei Perkov\,$^{7}$,
Sergei Bondar\,$^{7}$,
Tatyana Burlakova\,$^{1,4}$,
Sergei Fabrika\,$^{1}$,
Iosif Romanyuk\,$^{1}$
}

\def\Address{
$^{1}$Special Astrophysical Observatory, Russian Academy of Sciences, Nizhnij Arkhyz, Russia\\
$^{2}$Space Research Institute, Russian Academy of Sciences, Moscow, Russia\\
$^{3}$Mathematics and Mechanics Faculty, Saint Petersburg State University, Saint Petersburg, Russia\\
$^{4}$Federal State Budget Scientific Institution Crimean Astrophysical Observatory, Russian Academy of Sciences, Nauchny, Crimea\\
$^{5}$Moscow Institute of Physics and Technology, Dolgoprudnii, Russia\\
$^{6}$Institute of Physics, Kazan Federal University, Kazan, Russia\\
$^{7}$Research Corporation “Precision Systems and Instruments”, Moscow, Russia
}

\def\corrAuthor{Oleg Ya. Yakovlev}
\def\corrEmail{yko-v@ya.ru}

\begin{document}
\onecolumn
\firstpage{1}

\title[Exoplanet survey by SAO RAS]{Exoplanet two square degree survey with SAO RAS robotic facilities} 

\author[\firstAuthorLast ]{\Authors} 
\address{} 
\correspondance{} 

\extraAuth{}

\maketitle
\begin{abstract}

\section{}
We used the 0.5-m robotic telescopes located at the Special Astrophysical Observatory of the Russian Academy of Sciences for monitoring two square degrees of the sky with the aim of detecting new exoplanets. A dimming of the visible brightness is expected due to the exoplanets transiting their host stars. We analyzed about 25000 raw images of stars taken in the period between August 2020 and January 2021 and plotted the light curves for about 30000 stars on a half-year timescale. Five newly discovered exoplanet candidates are being investigated to determine their transit event parameters. We also present the light curves for dozens of binary stars. 

\tiny
 \keyFont{ \section{Keywords:} exoplanets, photometry, transit method, robotic telescope, variable stars}
\end{abstract}

\section{Introduction}

At present, exoplanets are detected and their parameters determined mainly by two indirect observation methods: photometric, where the light curve is analyzed when an exoplanet passes across the disk of the host star (transit method) \citep{Transit}, and spectral measurements, used for analyzing the radial velocity curve of the host star influenced by the gravitational pull of the exoplanet (RV method) \citep{RV}. In ground-based observations, most exoplanets (1296 out of 1494) \citep{NASA} were detected using the RV method (910) or the transit method (386). In space observations, the most used technique is the transit method (3387 out of 3426), while the RV method is not used (hereinafter, exoplanets and their host-star parameters and counts are taken from \citep{NASA} relevant for 2022.02.18).

\begin{figure}[h!]
	\begin{center}
		\includegraphics[width=85mm]{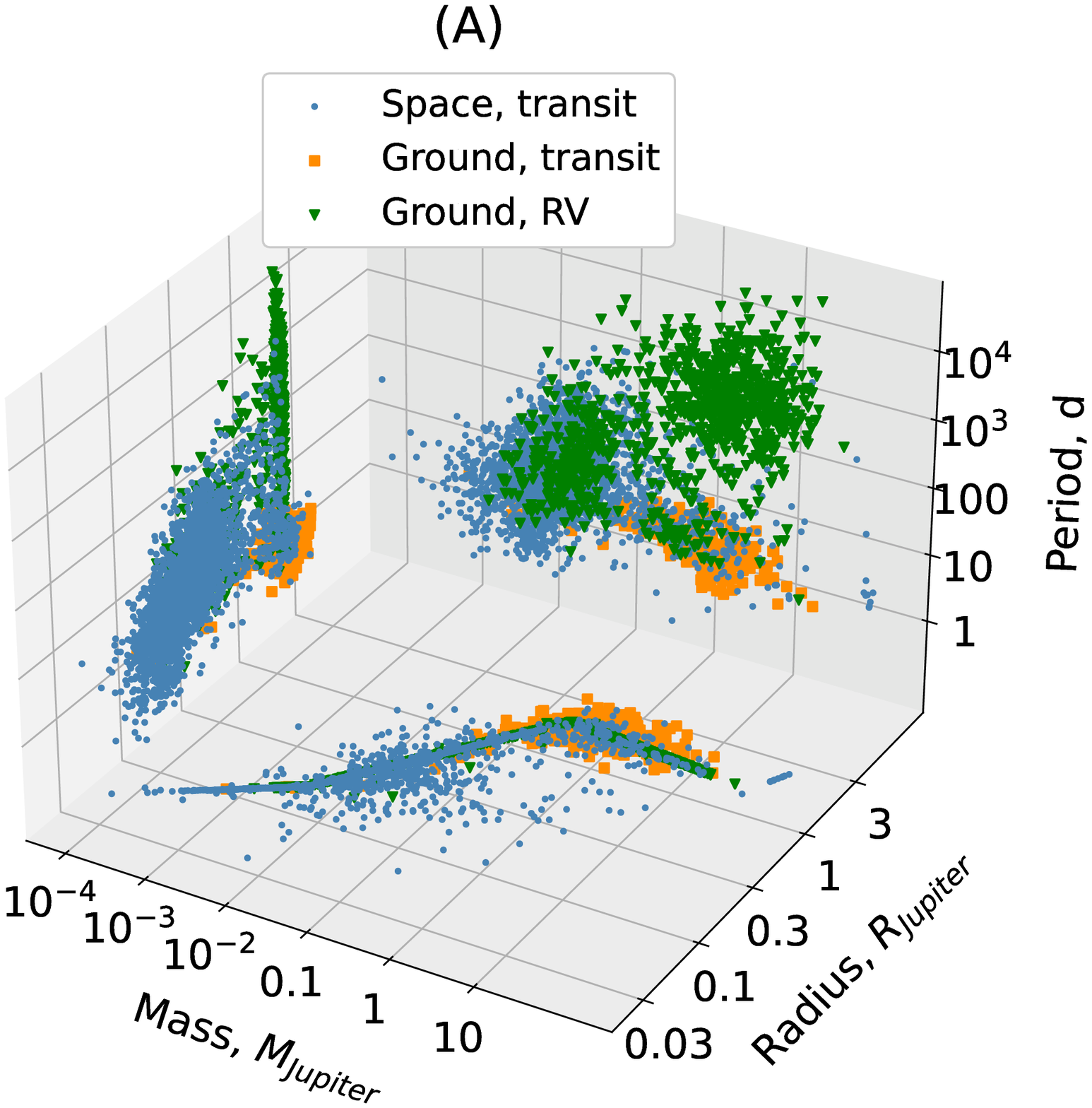}
		\includegraphics[width=85mm]{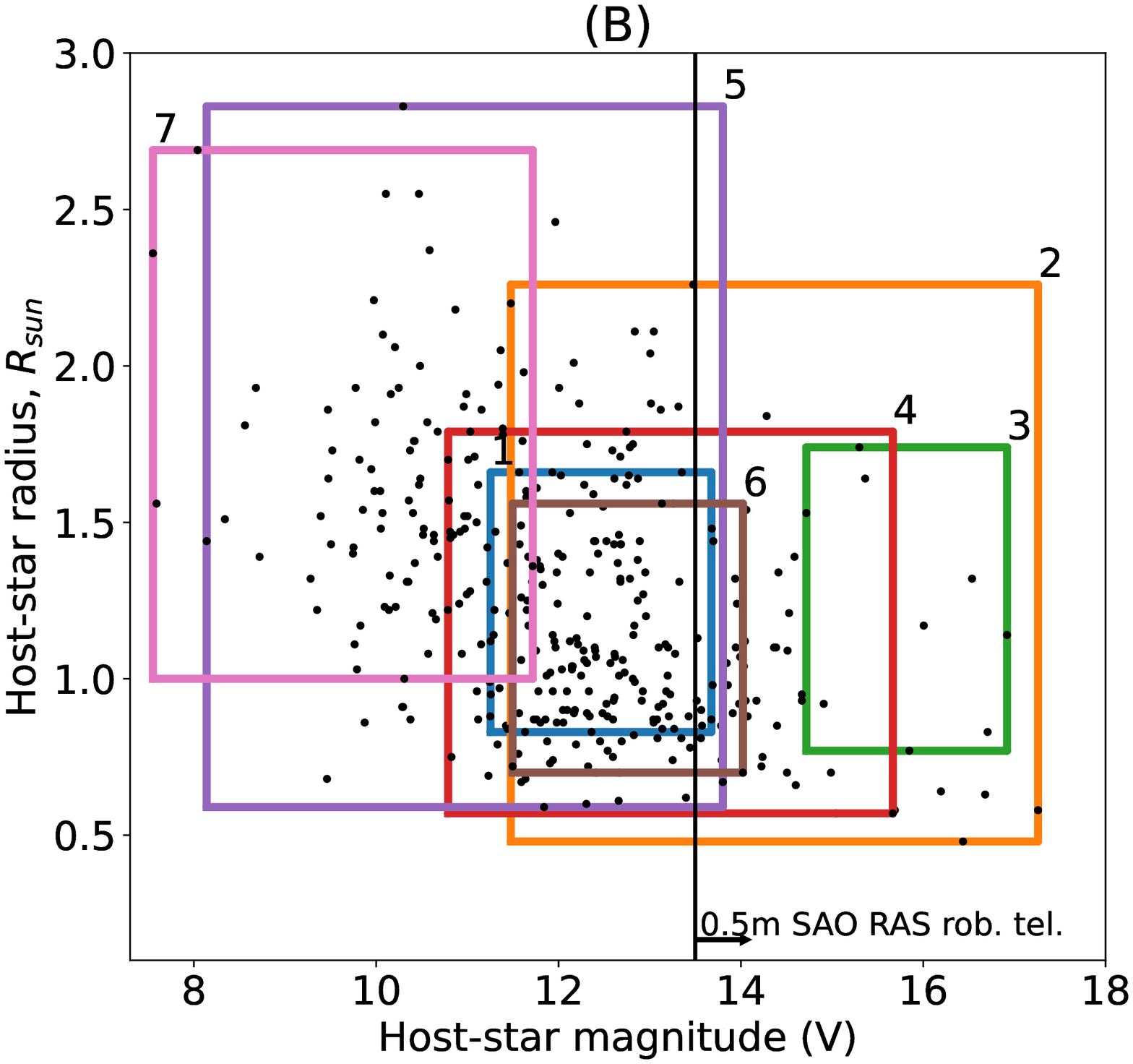}
	\end{center}
	\caption{\textbf{(A)} Confirmed exoplanets \citep{NASA} in the mass-radius-period planes depending on the discovery locale (space, ground-based) and on the detection method (transit, RV). For exoplanets with unknown mass or radius, they were calculated using the mass-radius model dependence. \textbf{(B)} Working areas in the host-stars magnitude-radius plane for 7 ground-based telescopes that detect exoplanets using the transit method (shown by dots, 347 out of 386): 1. 0.1m Schmidt Telescope, 2. 0.18m Takahashi Epsilon Astrograph, 3. 1.3m Warsaw University Telescope, 4. 0.2m Telescope, 5. 0.2m Canon, 6. 0.4m Canon, 7. 0.08m Mamiya 645. Also shown is the minimum limiting magnitude of the considered stars (${13.5^m}$) for data that were obtained with the 0.5-m SAO RAS robotic telescope and are discussed in this paper; the maximum limit (${19.5^m}$) is not shown.}
	\label{fig:1}
\end{figure}

The region in the exoplanet and host star parameter space in which exoplanets are detected is determined by the technical capabilities of observational instruments, the features of exoplanet detection methods, and the presence of an atmosphere in ground-based observations. Space telescopes (CoRoT \citep{CoRoT}, Kepler \citep{Kepler}, TESS \citep{TESS}) are more focused (Fig. \ref{fig:1}a) on detecting small and light exoplanets (Earth-like planets, mini-Neptunes), while ground-based telescopes (Super-WASP \citep{WASP}, HATNet \citep{HATNet}) detect mostly large and heavy ones (mainly gas giants). Most exoplanets with short orbital periods were detected by the transit method: those with periods of less than 50 days were observed by space telescopes, and those whose periods are less than 10 days, by ground-based telescopes. Exoplanets with long periods up to 40,000 days (106 years) \citep{Rosenthal} where detected by ground-based telescopes using the RV method.

Ground-based observations are weighed down by the Earth's atmosphere which produces star scintillation and variation of seeing, and the duty cycle of observations. The wavefront perturbations of stellar light passing through the atmosphere reduce the signal-to-noise ratio, which limits photometric precision and the minimum brightness dimming that can be registered. The brighter the star, the greater the signal-to-noise ratio, but the decrease in brightness during a transit will be less significant. Therefore, exoplanets near stars with magnitudes ${m=10–14^m}$ are mainly detected (300 out of 386) by the transit method in ground-based surveys (Fig. \ref{fig:1}b). The daily rotation and annual motion of the Earth determine the possibility of observing a given section of the sky from a given observation point. Therefore, the transit of an exoplanet may occur at those moments of time when it cannot be observed, several times in a row, although the technical capabilities make it possible to register the observed decrease in brightness. For reliable detection of exoplanets, one must observe at least several transit events that occur with a period equal to the exoplanet orbital period (for example, it takes several years or more to detect planets at a distance of more than 1 AU from a Sun-like star).

Thus, for the detection of exoplanets by the transit method in ground-based surveys, it is especially important to increase the number of observed stars and the observation time. For these purposes, it is relevant to use robotic telescopes that regularly perform long-term routine observations of a certain part of the sky (to increase the observation time) or different parts of the sky (to increase the number of stars). The use of a group of such telescopes makes it possible to achieve both goals.

A group of robotic telescopes is currently being developed at SAO RAS with the aim of detecting exoplanets by the transit method and conducting additional observations of the already known exoplanets. Currently, one such telescope is operating, automatically observing two regions of the sky since the summer of 2020, each for six months. We have developed software to perform photometric analysis and to search for transit events in the light curves of stars. In this paper we present the analysis and primary results of processing the first data set obtained in the last third of 2020.

The aim of this work is to examine the obtained data for suitability for the search for exoplanets. To that end, it was necessary to go through all the stages of image processing, determine the problems that arise in the process of data processing, and evaluate the possibility of using algorithms to search for transit events in the light curves. Section 2 is devoted to the description of the observations and the quality of the obtained data. Section 3 briefly describes the stages of the developed pipeline, and also discusses the problems that arise when working with the data. The primary results of the search for exoplanet candidates are presented in Section 4, which also presents the light curves for some of the found variable stars. A discussion of the five detected candidates and the prospects of our exoplanet survey are presented in section 5.

\section{DATA AND METHOD}
\subsection{Robotic telescope array}
The "Astrosib" (Novosibirsk, Russia) RC-500 telescopes with a 0.5-m  hyperbolic main mirror were installed on the  fast tracking "10 Micron GM 4000" high precision equatorial mounts. In the primary focus each unit was equipped  with an FLI Proline PL16801 front illuminated CCD  with a 9~$\mu$m pixel size. The Baader Planetarium AllSky 4.5-m dome with a weather forecast meteo station and a 4.5-m Astrosib AllSky Dome were used as  shelters. The FLI CCD camera,  FLI Atlas focuser and FLI five-position motorized 50-mm filter wheel are operated by an industrial PC which collects the raw data and provides remote access. 

The cloud and humidity sensors automatically send the telescope into parking position and signal for the dome to close. The twilight sky flat-field correction frames have been obtained since February 2021. Bias and dark calibration frames are collected daily.

\subsection{Observations}

\begin{figure}[h!]
	\begin{center}
		\includegraphics[width=10cm]{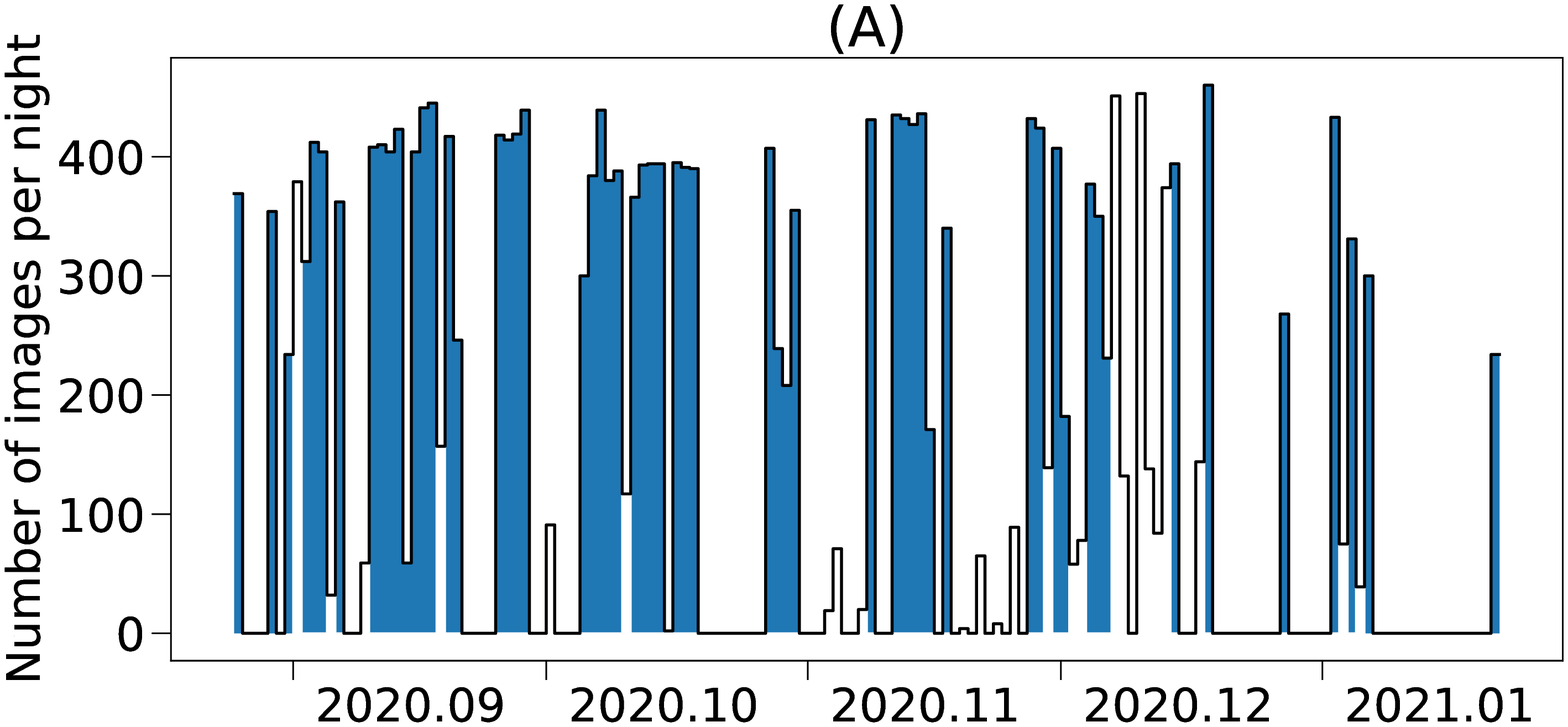}
		\includegraphics[width=6cm]{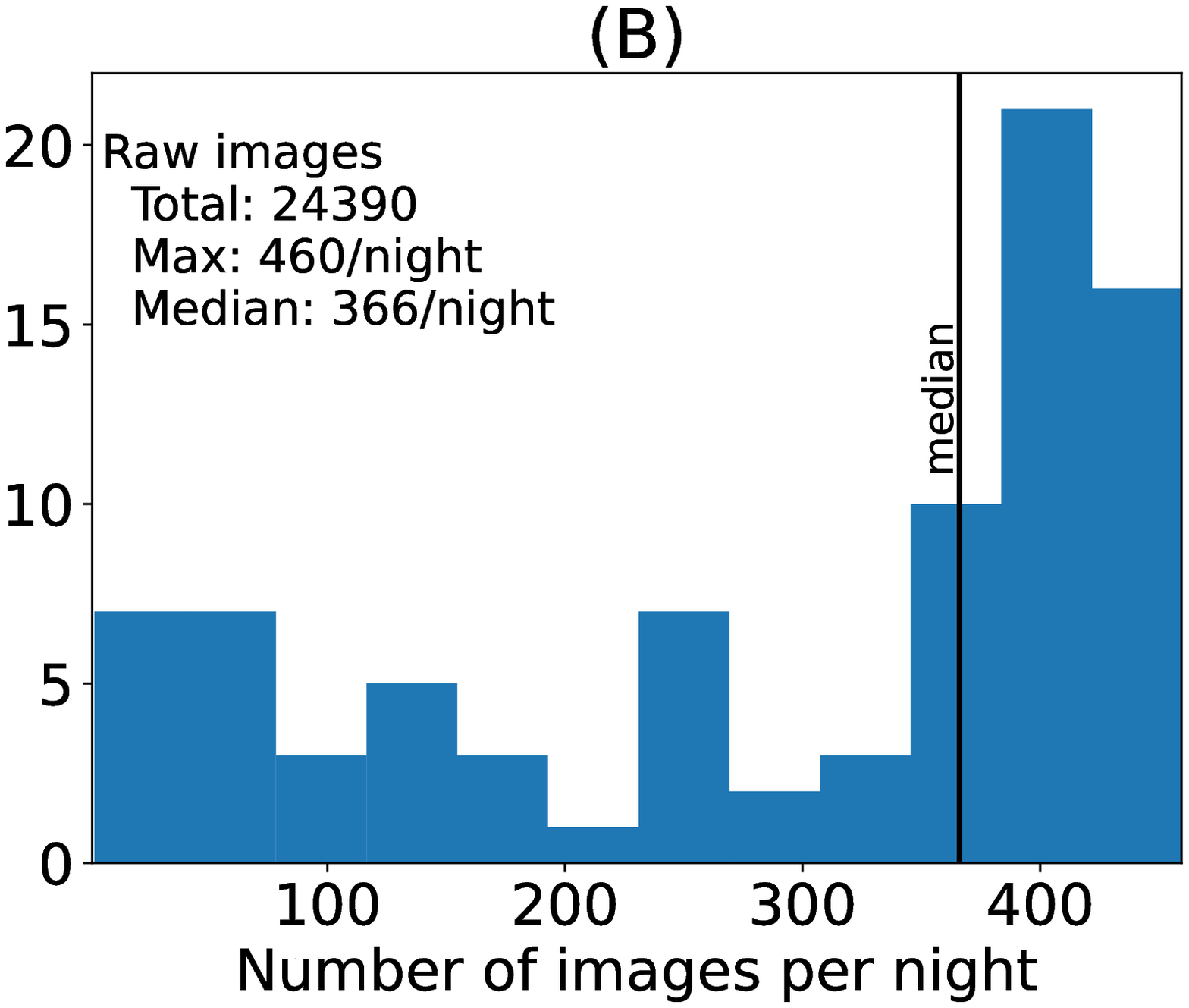}
	\end{center}
	\caption{Statistics on the number of images per night: \textbf{(A)} distribution by time with the colored columns marking the nights with successfully performed photometry, \textbf{(B)} distribution by number.}\label{fig:2}
\end{figure}

From August 25, 2020 to January 21, 2021, surveys were carried out for 84 nights (Fig. \ref{fig:2}a). Of these, 56 turned out to be suitable for light curve analysis; they are colored in Fig. \ref{fig:2}a. The white dwarf WD0009+501 with a well known 8-hour period and 5 mmag amplitude variations was chosen as the central object in the frame \citep{WD}. The ${2.45 \times{1.56}^\circ}$ (RA x DEC) field of view around this object has good visibility conditions in the autumn.

Observations were carried out with an exposure of 60 s and with a period of 80–100 s in the V filter of the Johnson system. The image scale is ${1.34''/pixel}$. Each night, 2 to 460 images were taken, the median value is 366. The total number of images is 24390 (Fig. \ref{fig:2}b). In addition, 10 dark and 10 bias frames were made for each night. No flat-field frames were taken.

\subsection{Data processing}
\begin{figure}[h!]
	\begin{center}
		\includegraphics[width=15cm]{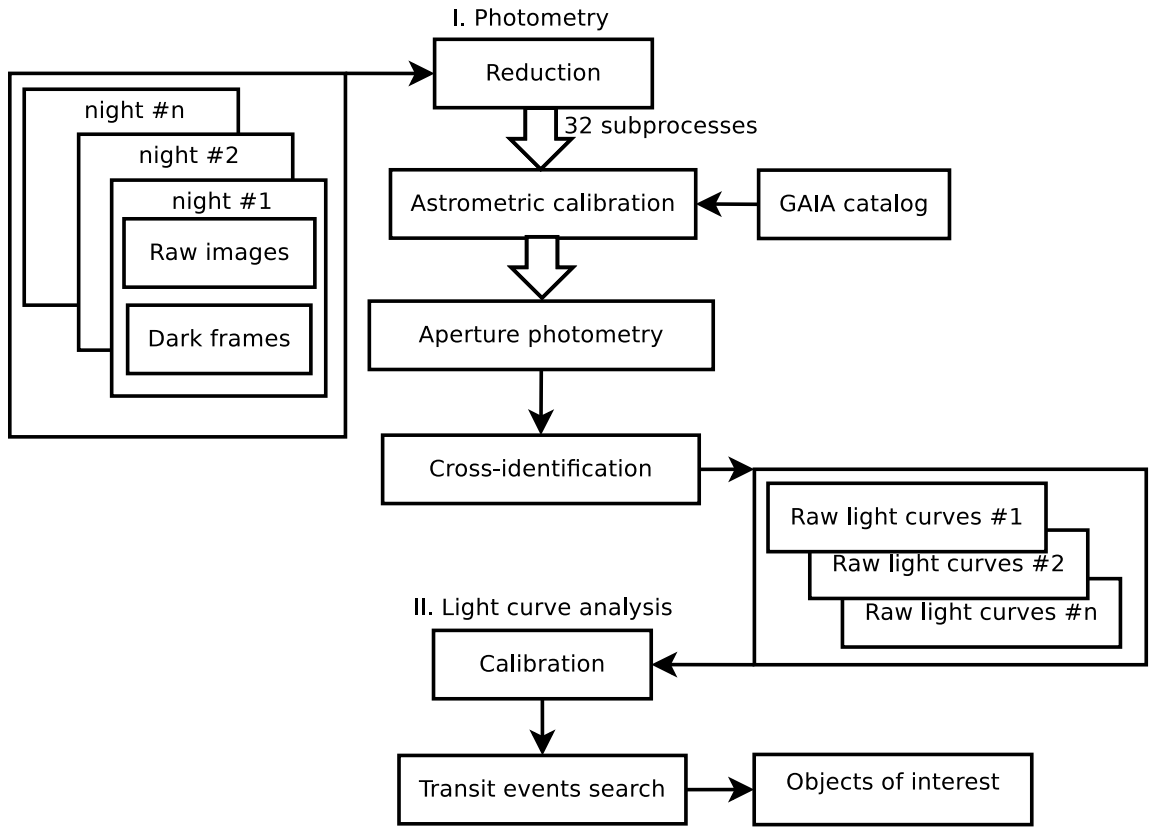}
	\end{center}
	\caption{Flowchart of the data processing algorithm for detecting exoplanets using the transit method.}\label{fig:3}
\end{figure}

The image processing is implemented in Python scripts combined into one bash-script. It can be divided into two independent parts (Fig. \ref{fig:3}): photometry – automatically obtaining light curves from images, and light curve analysis – searching for dimming in the light curves.

A catalog of stars in the considered field was prepared in advance according to the data of the GAIA space telescope (\cite{GAIA1}, \cite{GAIA2}): ID, coordinates (ra, dec), magnitude. This catalog contains 39978 stars with magnitudes ${m\in[13.5, 19.5]^m}$. 

\subsubsection{Photometry}

The entire field is divided into 9 overlapping parts: central (2500x2500 pix), 2 sides (820x2500 pix), top and bottom (2500x820 pix), 4 corners (1000x1000 pix). For each of the 9 parts of the frame, the processing is carried out sequentially for each night, which takes up to 5 minutes of machine time (CPU frequency 2.2 GHz). Obtaining light curves for one part of the frame for all nights takes from 2 to 4 hours. Reduction, astrometry calibration and aperture photometry are carried out in a multiprocessing mode, which makes it possible to process 32 images simultaneously at these stages.

The first stage of the pipeline is to create a configuration file based on the parameters entered by the user. After that, a calibration file is prepared. 10 dark frames are combined, the resulting dark frame is subtracted from the images taken at night. Further, using the astrometry.net package \citep{Astrometry}, the images are calibrated to the celestial coordinate system -- the coefficients of the transition matrix from the rectangular coordinates of the image x, y to astronomical equatorial coordinates ra, dec are determined. Reduction and astrometry calibration are carried out using the CCDpack package \citep{CCD}.

\begin{figure}[h!]
	\begin{center}
		\includegraphics[width=85mm]{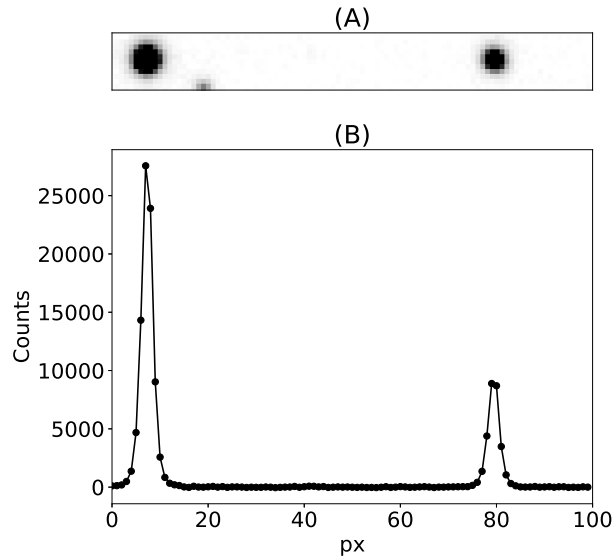}
	\end{center}
	\caption{An example with two stars in the image after reduction \textbf{(A)} and their projection onto one axis, where each point corresponds to the counts in a pixel \textbf{(B)}.}\label{fig:4}
\end{figure}

The next step is aperture photometry using the SExtractor package \citep{SExtractor} in 8 apertures with a diameter of 4 to 12 pixels. The PSF of stars falls within this range under various weather conditions (an example is shown in Fig. \ref{fig:4}). For identification of stars, the previously prepared GAIA catalog with an association radius of 5 pixels is used. In the resulting table each star from the catalog identified with the source in the image is assigned an instrumental magnitude for each aperture.

\begin{figure}[h!]
	\begin{center}
		\includegraphics[width=75mm]{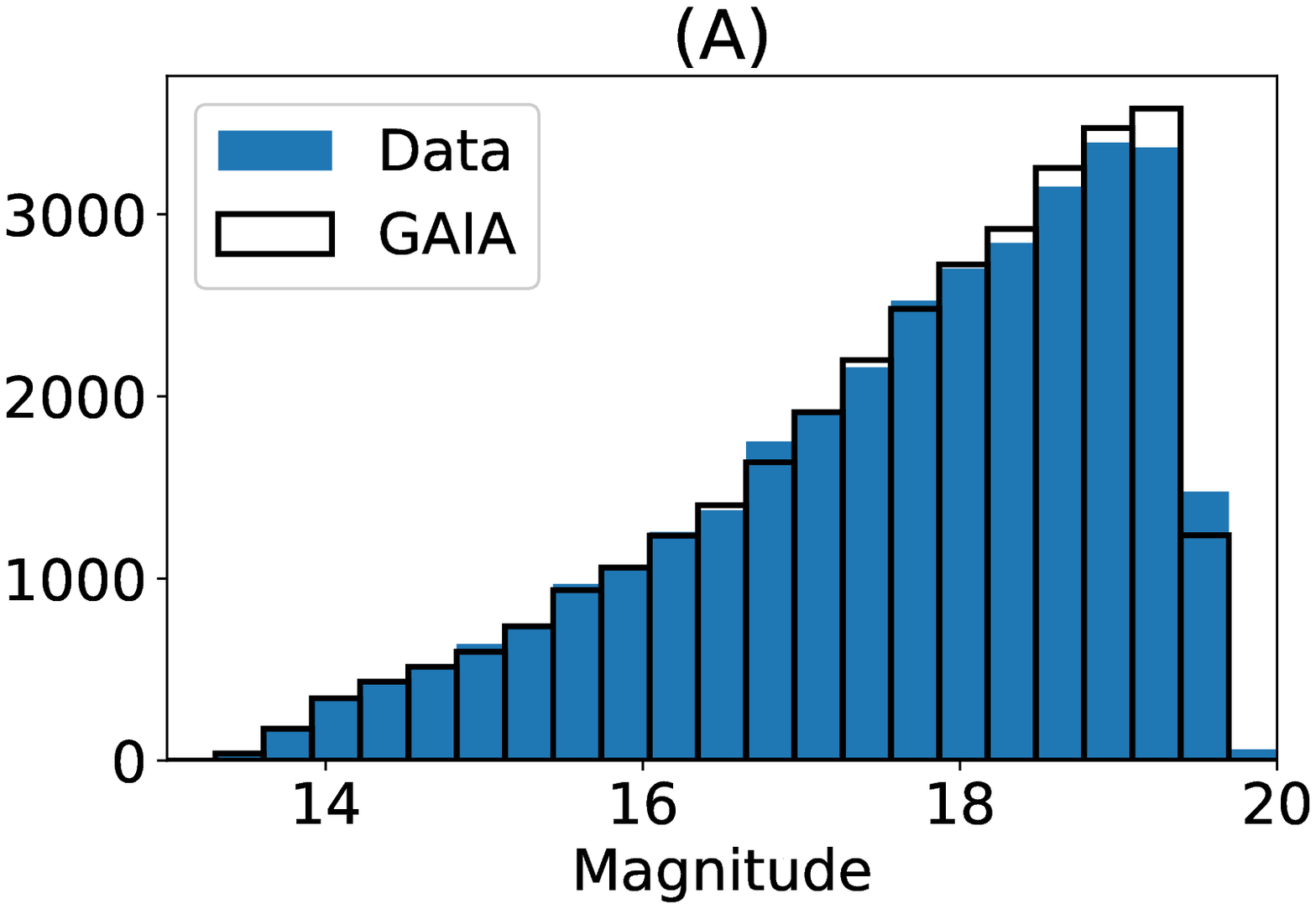}
		\includegraphics[width=95mm]{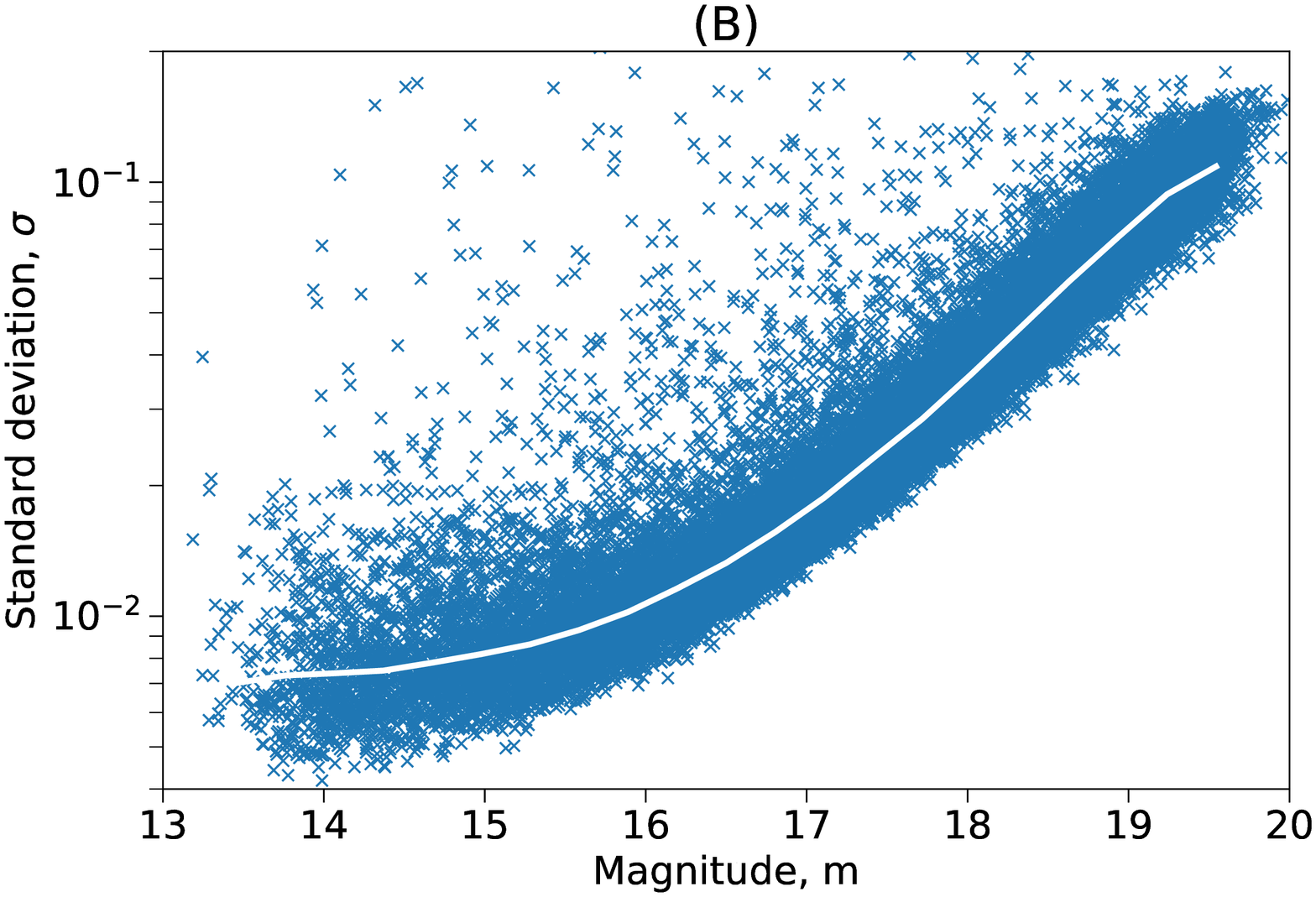}
	\end{center}
	\caption{Photometry statistics for one night: \textbf{(A)} magnitude distribution compared to GAIA catalog (\cite{GAIA1}, \cite{GAIA2}) and \textbf{(B)} standard deviation versus magnitude, with the median dependence shown by the white curve.}\label{fig:5}
\end{figure}

Further, the obtained tables for each image taken at night are combined into a single catalog, in which instrumental magnitudes in 8 apertures for the entire night are assigned to each identified star. Thus, as a result of performing the described steps, the light curves for the stars from the input catalog are obtained from raw images for each night. Fig. \ref{fig:5}a shows a magnitude distribution example for the stars according to the catalog and derived from photometry. For some stars with magnitudes of ${m=14^m}$, the standard deviation reaches minimum values up to ${0.005^m}$ (Fig. \ref{fig:5}b).

To select the operating range of magnitudes, we first analyzed the dependence of magnitude, according to the GAIA catalog, on the number of counts on the CCD array. ${13.5^m}$ was taken as the lower limit, ${19.5^m}$ as the upper limit. In this range, there are no stars with saturated pixels or those whose signal levels are comparable with the noise levels. 

\subsubsection{Light curve analysis}
\begin{figure}[h!]
	\begin{center}
		\includegraphics[width=18cm]{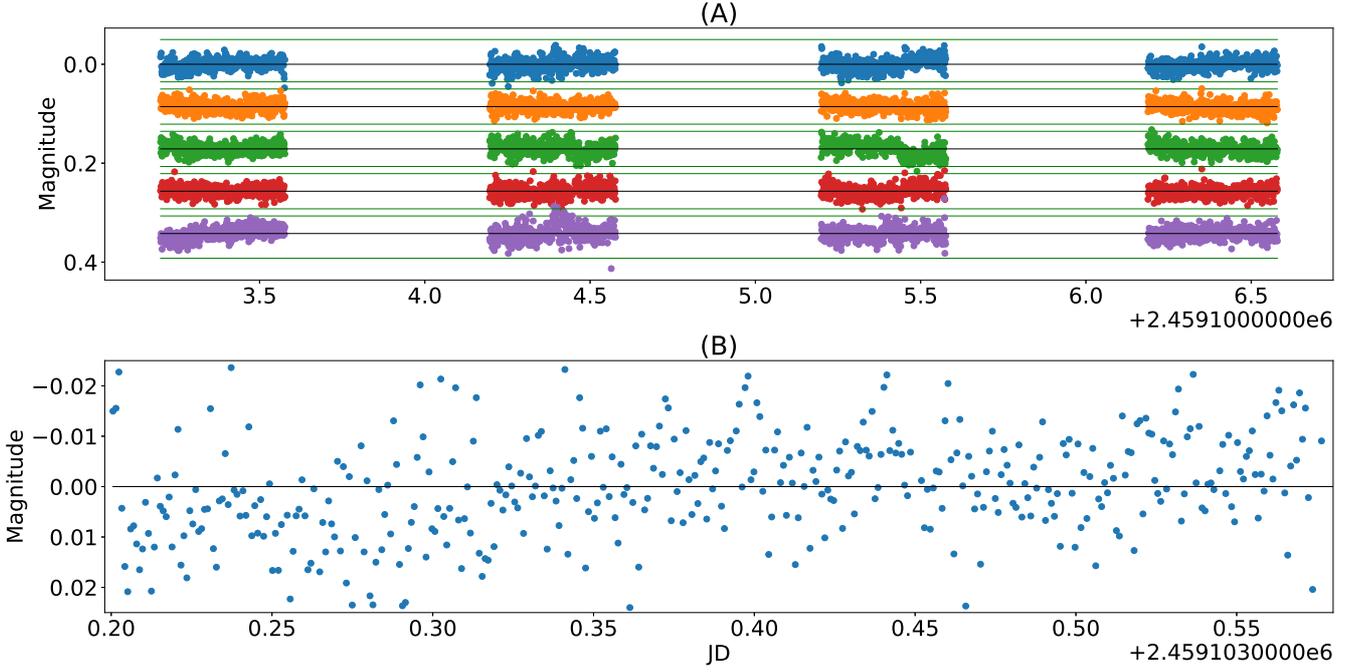}
	\end{center}
	\caption{Light curve examples for 4 stars with ${m\approx15.5^m}$: \textbf{(A)} for 4 nights, with a ${\pm0.05^m}$ zone shown for each star, \textbf{(B)} for the first star on the first night.}\label{fig:6}
\end{figure}

The light curves obtained for each night are combined into one for all nights. After that, they are calibrated. Stars are selected as standard stars if they satisfy the following conditions: i) they are identified on all nights, ii) their magnitudes are less than 15.5, iii) they are in the same part of the frame as the target star. The light curves for several standard stars are shown in Fig. \ref{fig:6}.

The last step is to search for transit events in the light curves using the BLS (Box Least Squares) method \citep{BLS}. In this method, the phase folded light curve is approximated by a two-level model: one level (high) before and after the transit, another level (low) during the transit. The function changes instantly at the beginning and at the end of the transit and is similar in shape to a box. The idea of the method is to calculate statistics depending on the difference between the model and the available data. The larger this statistic, the smaller the residual. For each trial period, the best combination of transit duration and transit start time is determined, which corresponds to the maximum value of the statistics. Thus, a periodogram is constructed -- the dependence of this statistic on the trial period. On the periodogram, a star with a light curve that exhibits a transit event has a global maximum, which differs significantly from the rest of the local maxima.
For each star, a periodogram is plotted for the light curves constructed using 5 standard stars. The threshold value at which the flag is triggered is set. For the most interesting objects, the flag is triggered twice or more. After that, the light curves of the selected objects are viewed individually to decide whether to consider the star as an object of interest.

\section{RESULTS}

\begin{figure}[h!]
	\begin{center}
		\includegraphics[width=12cm]{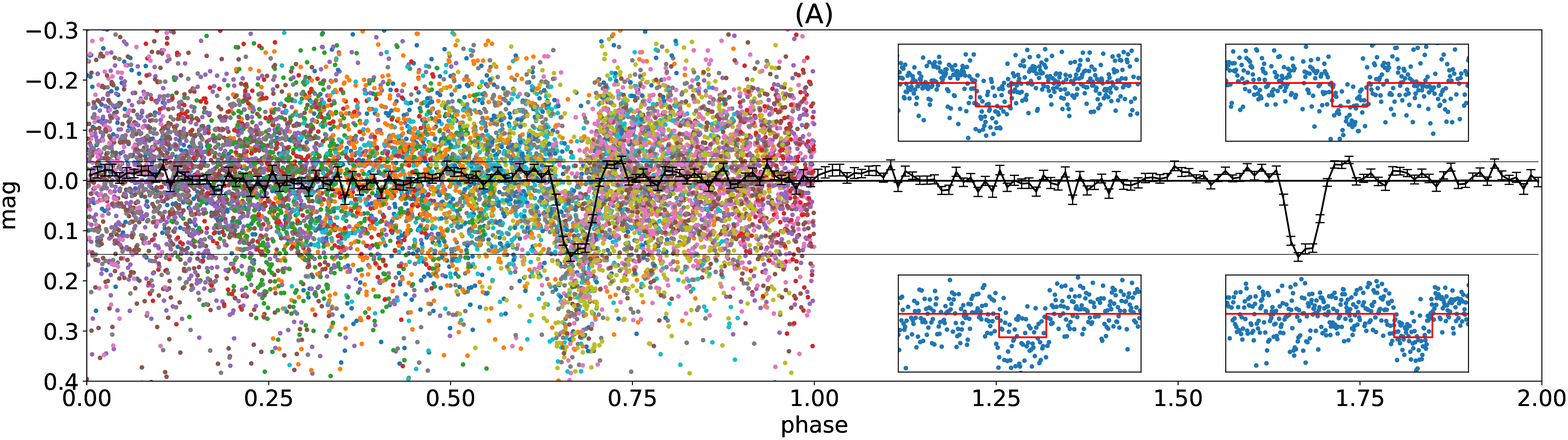} \\
		\includegraphics[width=12cm]{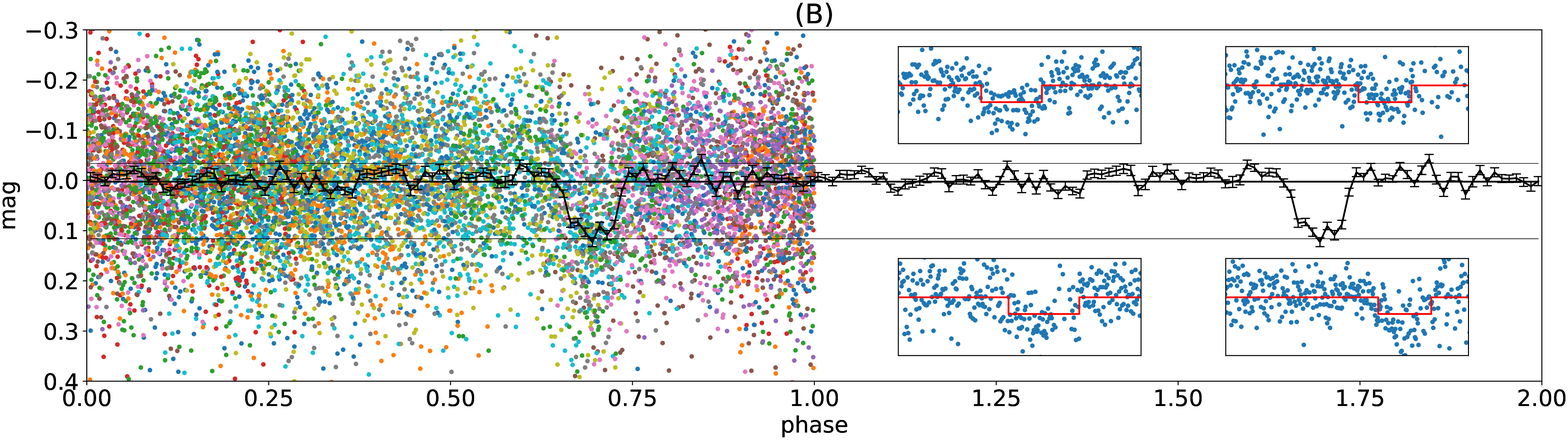} \\
		\includegraphics[width=12cm]{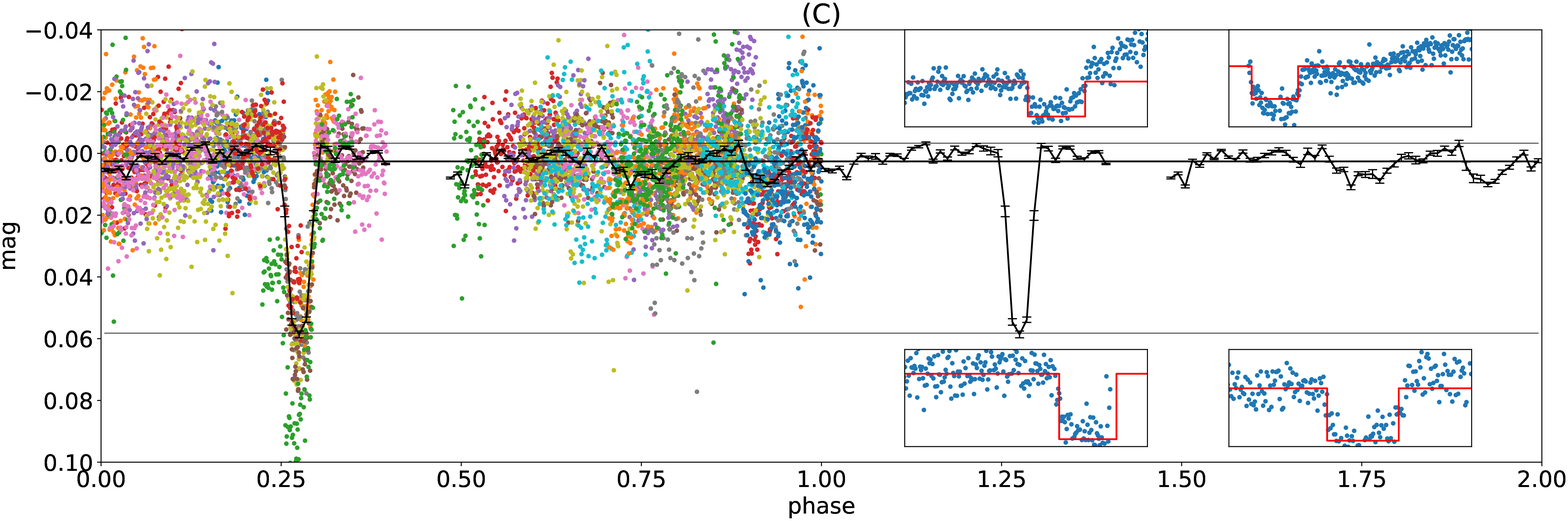} \\
		\includegraphics[width=12cm]{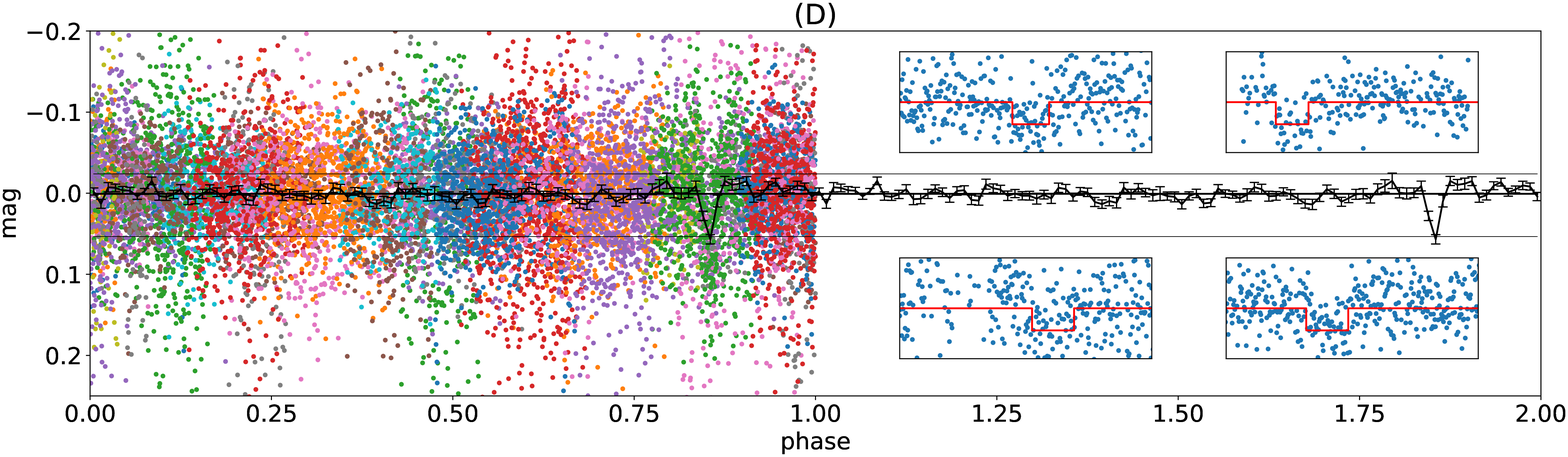} \\
		\includegraphics[width=12cm]{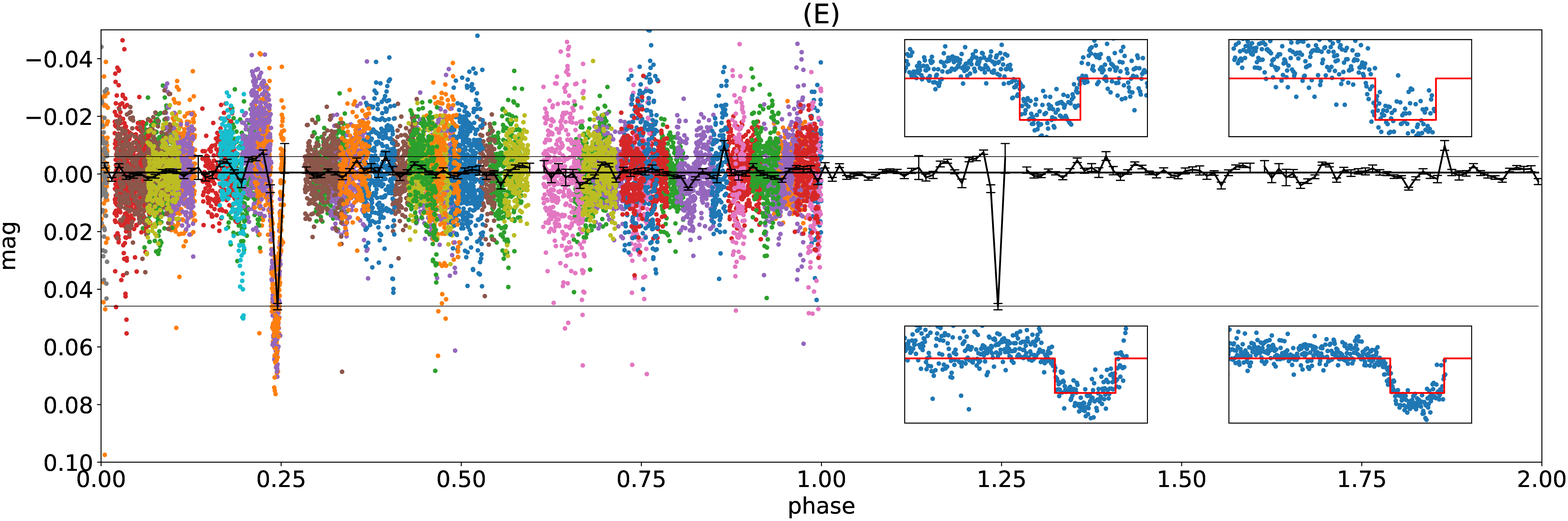}
	\end{center}
	\caption{Folded light curves of five objects of interest and binned averaged curves  (main panels). Points from different periods are shown in different colors. The horizontal lines show the minimum, maximum and median values. The four small panels show examples of individual transit events with the model found by the BLS method (see their parameters in table \ref{tab:1}).}\label{fig:7}
\end{figure}

As a result of applying the algorithm to the obtained data (Section 2), five interesting objects were found (named SOI – SAO RAS Objects of Interest). Their light curves and parameters are shown in Fig. \ref{fig:7} and table \ref{tab:1}, respectively.

\begin{table}[h!]
	\begin{center}
	\begin{tabular}{ c | ccccc}	
		Object 			& SOI-1 & SOI-2 & SOI-3 & SOI-4 & SOI-5 \\
		\hline
		Magnitude, m 	& 18.8 & 18.9 & 14.3 &	17.8 &	14.9 \\
		Period, h 		& 26.1 & 25.2 & 46.0 &	63.4 &	198.3 \\
		Depth, m 		& 0.1 &  0.07 & 0.05 &	0.04 &	0.04 \\
		Duration, h 	& 1.4 & 2.0 & 1.7 &	1.6 &	2.4 \\
	\end{tabular}
	\end{center}
	\caption{Objects of interest and transit event parameters.}
	\label{tab:1}
\end{table}

\begin{figure}[h!]
	\begin{center}
		\includegraphics[width=15cm]{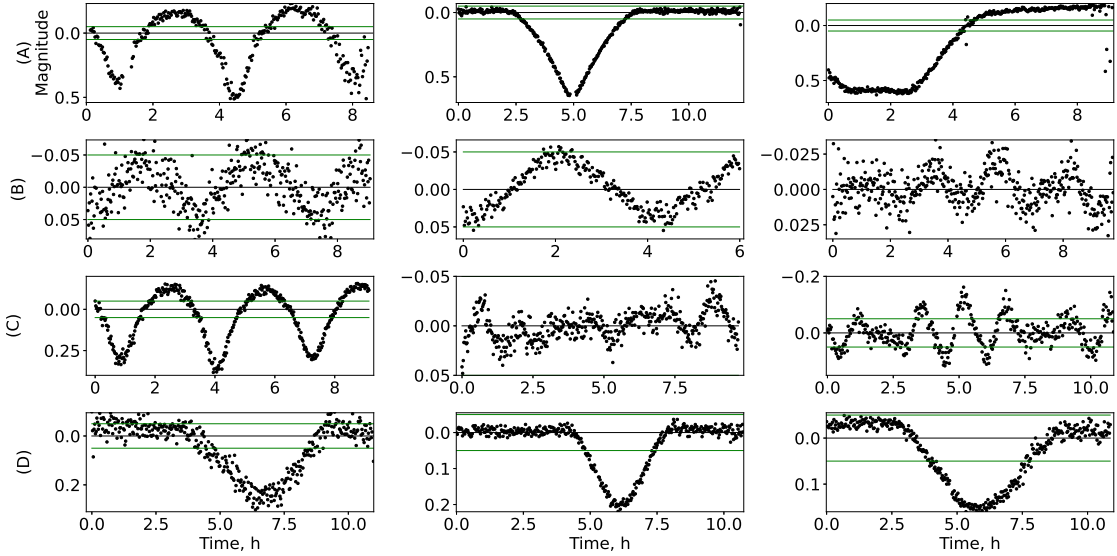}
	\end{center}
	\caption{Light curves of variable stars: \textbf{(A)} with large amplitudes (${\Delta{m}>0.5^m}$), \textbf{(B)} with small amplitudes (${\Delta{m}<=0.1^m}$), \textbf{(C)} with short periods (${P<4^h}$),  \textbf{(D)} with a shape similar to a transit event. The horizontal lines show the ${\pm0.05^m}$ zone.}\label{fig:8}
\end{figure}

As a result of the deviation-magnitude relation analysis (Fig. \ref{fig:5}b), as well as in the process of searching for exoplanets, more than 90 variable stars were found; the light curves for 12 of them are shown in Fig. \ref{fig:8}.
A detailed description of the selection criteria for stars in multiple systems or stars with exoplanet candidates will be given in the next paper. The validation of their binary nature (multiplicity) using additional spectral data from the 6-m BTA telescope has just started.

\section{DISCUSSION}

\begin{figure}[h!]
	\begin{center}
		\includegraphics[width=8cm]{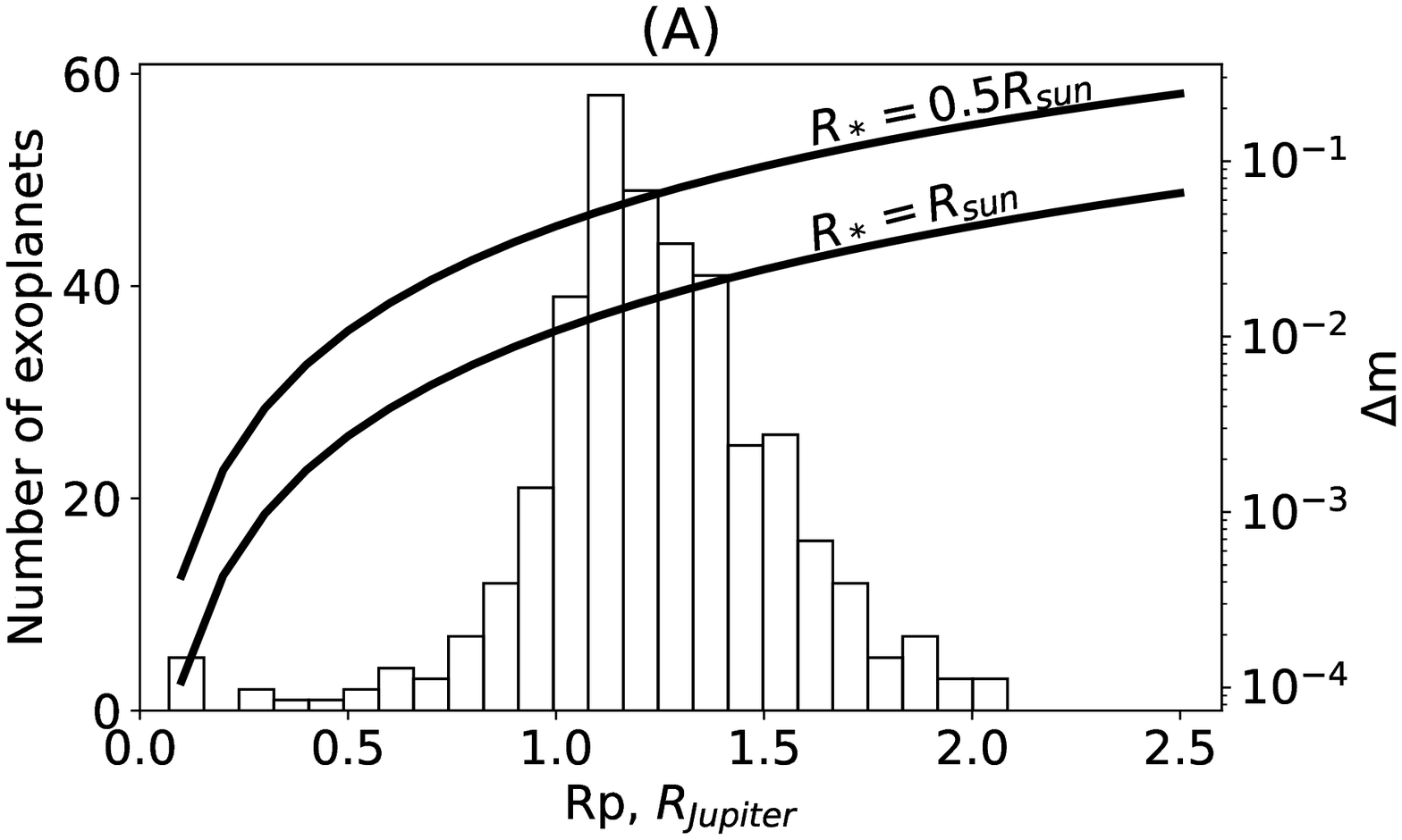}
		\includegraphics[width=9cm]{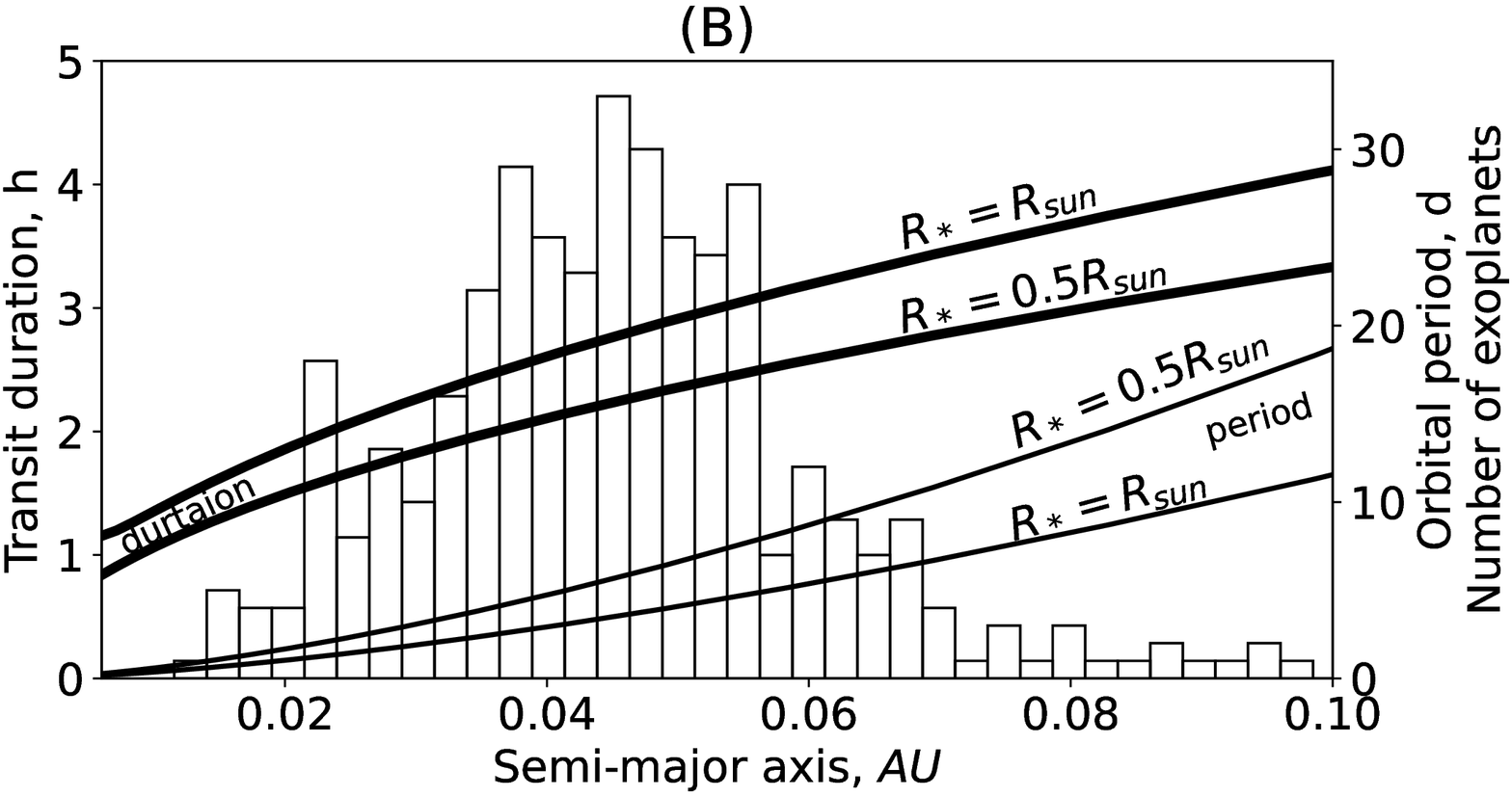}
	\end{center}
	\caption{Maximum brightness dimming of the host-star during a transit event ${\Delta{m}}$ versus exoplanet radius (\textbf{(A)}, right axis) and \textbf{(B)} transit duration (left axis, bold lines) and orbital period (right axis, thin lines) versus semi-major axis of the exoplanet orbit for stars of solar radius and those two times smaller. Also shown are the histograms of the distribution of confirmed transiting exoplanets \citep{NASA} detected by ground-based facilities by exoplanet radius (\textbf{(A)}, left axis), and by the semi-major axis of their orbits (\textbf{(B)}, right axis).}\label{fig:9}
\end{figure}

We have shown in practice that exoplanet candidates can be found from the obtained data. The dimming in the objects of interest is in the range of ${\delta m\approx{0.04...0.1^m}}$, which corresponds to a transit of a 1-2 Jupiter radius exoplanet across the disk of a star of solar radius or less (Fig. \ref{fig:9}a). They have an orbital period of 1-8 days, corresponding to a planet orbiting close to the host-star, with the semi-major axis less than about 0.08 AU (Fig. \ref{fig:9}b). These estimates indicate that hot Jupiters or, possibly, hot Neptunes are the targets of our search for exoplanet candidates. After processing the data in this sky area for the second half of 2021, we plan to search for exoplanets in wider orbits with periods of up to several months. 
The transit durations of all the SOI with determined periods are less than the corresponding maximum value; they pass this initial test successfully. SOI-2 remains the most unclear.
The odd and even minima in the light curve of SOI-2 differ.
Furthermore, based on the detected objects, we can improve the algorithm by refining the criterion for selecting transit events.

The detected objects must undergo further validation. To that end, we plan to carry out additional photometric and spectral measurements. The light curves in different filters and the radial velocity (RV) amplitudes of the stars will allow us to reject the double star hypothesis. The shape of the residual between the light curves in different filters during the transit of an exoplanet differs from the shape for an eclipsing binary star with different spectral type components. The RV amplitude determines the minimum mass of the component that produces dimming.
On the one hand, most ground-based surveys observe brighter stars, so our observations in the range of ${m\in[13.5, 19.5]^m}$ can be productive. However, the necessary validation of faint objects by the RV method is more difficult. There is also less information about faint stars, so modeling transit events for validation purposes is more uncertain. Therefore, it is possible that in the future this range will be shifted towards brighter stars.

The disadvantage of the existing pipeline is the different orientation of raw images, which complicates the last step of creating the light curve, when the star's light curves converge into one for all nights. Therefore in the future we plan to transform each image at the first stage, bringing all the images to the same orientation. In addition, it is possible to speed up the process by forming packages of consecutive images to reduce the time of astrometric calibration. At the light curve analysis stage, detrending needs to be added.

Currently, the data have been processed only for the last third of 2020, but the same sky field was observed in 2021; 2021 data are also available for another field. After processing these data and improving the algorithm, the pipeline will be introduced into the observing process. Processing will be carried out immediately after the observations.

\section*{Funding}
The study was supported by the Government and the Ministry of Education and Science of Russia (grant no.075-15-2020-780 (N13.1902.21.0039)). VAF acknowledges the star multiplicity advocation algorithms development was   supported by the Russian Science Foundation grant 19-72-10023.

\section*{Acknowledgments}
This work has made use of data from the European Space Agency (ESA) mission {\it Gaia} (\url{https://www.cosmos.esa.int/gaia}), processed by the {\it Gaia} Data Processing and Analysis Consortium (DPAC, \url{https://www.cosmos.esa.int/web/gaia/dpac/consortium}). Funding for the DPAC has been provided by national institutions, in particular the institutions participating in the {\it Gaia} Multilateral Agreement.

In this paper, we used the NASA Exoplanet Archive, which is operated by the California Institute of Technology, under contract with the National Aeronautics and Space Administration under the Exoplanet Exploration Program.

\bibliographystyle{Frontiers-Harvard}
\bibliography{frontiers}

\end{document}